# The Impact of e-Politician on the Adoption of e-Service: Perceptions from a sample of South African Municipal IT Heads

Ntjatji Gosebo[1]
[1]Department of Public Service and Administration, South Africa

## Abstract

*The purpose of this study is to establish whether the use of information technology (IT) by elected municipal representatives, for constituency work, emboldens the adoption of e-service in municipals of a developing country. The research data was obtained through the completion of a questionnaire by a sample of respondents who serve as authorities of IT in South African municipals. The findings from both descriptive and inferential data analysis of collected data confirm that the use of IT by elected municipal representatives for constituency work impacts the adoption of e-service in municipals. Furthermore, the use of IT by elected municipal representatives for constituency work correlated with both e-service laws and e-service security. This study contributes to a better understanding of choices needed when planning for the adoption of e-service initiatives in municipals of developing countries. Given that 87.2% of respondents are aware of a high access to telephone mobile, a further research is needed to clarify why most elected municipal representatives of a developing country choose not to exploit IT for their constituency work, and similarly why municipals of a developing country do not exploit IT to provide services.*

## Keywords

*e-politician, e-service adoption, municipal, developing country, elected municipal representative*

## 1. Introduction

Elected government representatives (politicians) use IT as a method to enhance their constituency work (Effing *et al*., 2011; Grant *et al*., 2010). Politicians with higher Social Media engagement got relatively more votes within most political parties in the 2010 national elections of Netherlands (Effing *et al*., 2011). The finding by Grant *et al*. (2010) also affirms that Australian politicians who used Twitter to converse appeared to gain more political benefit from the platform than others. Therefore, to what extent does the use of IT for constituency work by elected municipal representatives of a developing country, impact the adoption of e-service in their respective municipals?

The concept of e-government has been described as a composition of e-organization, e-service, e-democracy, and e-politician (Carrizales, 2008; Serrano-Cinca *et al*., 2009; Kabir and Baniamin, 2011; Weerakkody *et al*., 2011). E-organization is mainly concerned with the use of information technology by governments for the optimization of their internal efficiency and effectiveness. E-service is defined as the use of information technology in the efficient and effective provision of public services to citizens. E-democracy consists in the use of information technology to ensure that citizens' voices are heard in public decision-making processes such as elections, petitions, etc. And e-politician simply refers to the use of IT by politicians in their political duties.

                                                                                    

Developing countries aspire to provide e-service, but, most of them lack sufficient levels of critical IT resources. Governments of developing countries do not necessarily have fully functional and integrated back-offices or infrastructure to support web services (Kamal, 2006). The lack of IT infrastructure is attested by the United Nations' e-government index of 0.2642 on Africa, which is below the average world e-government index of 0.4267 (Imran and Gregor, 2007). Thus, e-service adoption for developing countries should start from establishing IT infrastructure, and thereafter ensuring peoples' access to IT infrastructure (Kabir and Baniamin, 2011; Gosebo and Eyono-Obono, 2012).

This study borrows the abovementioned e-service adoption concept as articulated by Gosebo and Eyono-Obono (2012) and Kabir and Baniamin (2011) for developing countries, to achieve the aim of the study. The municipal IT head is considered the best informant on the IT usage by politicians, because overall IT support and maintenance in organizations is the IT head's responsibility (Carrizales, 2008; Li and Qiu, 2010). Therefore, this study aims to analyse whether IT usage by elected municipal politicians for constituency work encourages the adoption of e-service, using lenses of municipal IT heads. Irani *et al*. (2012) found that e-government studies conducted between 2002 and 2012 used government IT and general staff as respondents in only 7.70% of all studies, whence the choice of this study's respondents as Municipal IT heads becomes important.

## 2. LITERATURE REVIEW

Various studies have established the importance that the support of elected government representatives (politicians) holds for the adoption of e-government (Arduini *et al*., 2010; Bwalya, 2009; Serrano-Cinca *et al*., 2008; Weerakkody *et al*., 2010). E-government initiatives are long term projects and need long-term financial support from the elected government representatives (Weerakkody *et al*., 2010). Political influences have been a major force in selecting and implementing information technology projects in government (Serrano-Cinca *et al*., 2008). Political support is reflected by technology adoption and decisions to undertake costly innovation efforts in the public sector (Arduini *et al*., 2010; Bwalya, 2009).

The support that politicians give to the adoption of e-services depends on how best e-service could advance objectives of their governments. Benefits of e-service to government objectives influence the adoption of e-service (Kamal, 2006; Potnis and Pardo, 2011). Other studies (Carter and Weerakkody, 2008; Rokhman, 2011) use the concept of relative advantage to imply benefits of e-service to government objectives.

Politicians are responsible for providing legislation to facilitate e-government legitimacy, with laws such as user authentication and e-government security. Basu (2004) submits that e-government friendly legislative revisions and enactment are required, as traditional laws, rules, and regulations might not recognize the legality of e-government transactions. It is not possible to transfer e-commerce solutions and development approaches directly to the public administration, because, of the legal framework that governs public administration (Alpar and Sebastian, 2005).

The ultimate approval for the municipal institutional arrangements to facilitate e-government rests with elected municipal representatives, and institutional arrangements as a factor for enabling e-government are recognized in literature (Al-Awadhi and Morris, 2009; Davison *et al*., 2005; Ebrahim and Irani, 2005; Elsheikh *et al*., 2008; Hassan *et al*., 2011). Reforming bureaucracy so as to smoothly facilitate e-government adoption was identified by studies that included Elsheikh *et al*. (2008) and Al-Awadhi and Morris (2009). Providing an enabling organization to facilitate e-government was recognized by studies that included Davison *et al.*, (2005), Ebrahim and Irani (2005), and Hassan *et al.* (2011).

Leadership of e-government, as an aspect of the preceding institutional arrangements, is critical for e-government adoption. Even cases where IT managers initiate the adoption of new technology, support from administrative authorities may play a significant role in whether innovation efforts are adopted (Kamal, 2006). The probability of e-government adoption increases with the presence and leadership of an IT department within a municipality (Carrizales, 2008). IT leadership of top managers is often directly associated with the inclusion of the IT head in the executive team (Li and Qiu, 2010).

The figure below encapsulates the foregoing presentation of what has been established in literature about the influence of the e-politician on the adoption of e-government.

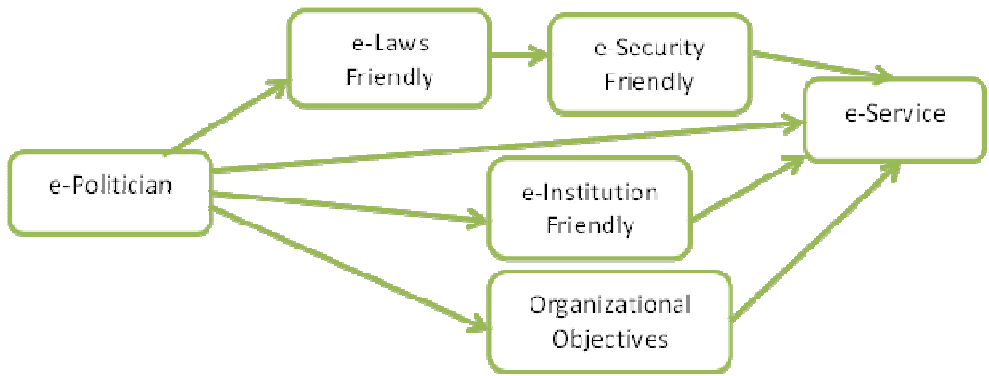

Figure 1 – Summary of Literature Review

Anyhow, none of these studies analysed the direct and indirect impact of e-politician on e-services as summarized in figure 2 from literature review of this study. Also, IT habits of a politician that impact e-service adoption were not a focus of any of the abovementioned studies. This paper aims to test whether the habit of using IT by municipal politicians for their constituency work influence the adoption of e-service in municipals of a developing country.

## 3. RESEARCH AIM AND THEORETICAL FRAMEWORK

The purpose of this paper is to analyse whether the use of IT by elected municipal representatives, in their constituency work, impacts the adoption of e-service in municipals of a developing country. This study is conceptually grounded in Rogers's model of the diffusion of innovations (DoI) theory, where Rogers (2003, 12) defines an innovation as "an idea, practice, or object that is perceived as new by an individual or other unit of adoption". Additionally, Rogers (2003:473) defines adoption as: a decision to make full use of an innovation as the best course of action available.

Rogers (2003) defines several intrinsic characteristics of innovations that influence an individual's decision to adopt or reject an innovation, as shown in table 1 below.

Table 1 - Innovation Characteristics (Rogers, 2003)

| Characteristic | Description |
|---|---|
| Relative advantage | How improved an innovation is over the previous generation? |
| Compatibility | The level of compatibility that an innovation has to be assimilated into an individual's life. |
| Complexity or Simplicity | If the innovation is perceived as complicated or difficult to use, an individual is unlikely to adopt it. |
| Trialability | How easily an innovation may be experimented. If a user is able to test an innovation, the individual will be more likely to adopt it. |
| Observability | The extent that an innovation is visible to others. An innovation that is more visible will drive communication among the individual's peers and personal networks and will in turn create more positive or negative reactions. |

The innovation process in organizations is much more complex. It generally involves a number of individuals, perhaps including both supporters and opponents of the new idea, each of whom plays a role in the innovation-decision.

## 4. RESEARCH DESIGN

This section describes how data was collected through a perceptions' survey instrument designed along the literature review summary of figure 1 above, while section 5 explains quantitative and inferential analysis of collected data using the SPSS software package.

### 4.1. Data Collection

Municipal IT heads were polled from all 9 provinces of South Africa. South Africa has 278 municipalities spread across 9 provinces. A choice to survey heads of municipal IT was made on the basis that they are expected to possess accurate and best information on the use of IT by politicians, also possess information about the levels of e-service in their respective municipals.
Data was collected in the form of an online questionnaire that ensured voluntary, anonymous, and convenient participation. The online questionnaire comprised of 65 items on the e-government adoption factors associated with politicians as identified earlier in the literature review and depicted in figure 1. The first research construct on the profile of the municipalities and of their heads of IT contained 15 items; while the remaining 5 research constructs were 5-point Likert-scale items with 10 items each. Among the research design, methodologies and approaches adopted in the extant e-government research studies, the quantitative research based approach supported by statistical analysis was the most dominant approach applied by authors in the last decade (Irani *et al*., 2012).

## 5. RESEARCH RESULTS

Data was quantitatively analysed using the SPSS software package. Reliability and validity tests were performed on collected data, followed by the descriptive and inferential statistical analysis of the collected data.

## 5.1. Data Reliability and Validity

Reliability and validity tests performed on the collected data found that research constructs of this study were reliable and valid for all the Likert-scale based research variables, as summarized in the table below.

Table 2 – Reliability and Validity of Likert-Scale Based Research Variables

| Construct | Survey items | Reliable Items | Valid Cronbach coefficient ($\alpha$) |
|---|---|---|---|
| IT security risks | 10 | 4 | 0.870 |
| Support for Municipal Objectives | 10 | 8 | 0.934 |
| Adequacy of the e-government regulatory framework | 10 | 10 | 0.894 |
| IT Usage by politicians | 10 | 6 | 0.894 |
| Adoption of e-government (resultant) | 10 | 10 | 0.915 |

The overall validity of the above Likert-scale based research variables is 0.915.

## 5.2. Descriptive Statistics

The analysis of the descriptive statistics computed by this study shows interesting results on the profile of municipalities and of respective heads of IT; and on general trends followed by the Likert-scale based research variables.

### 5.2.1. A Profile of municipality and its head of IT

The spread of responses received from municipal IT heads ranged between 5.77% and 19.23% across 9 provinces of South Africa, whence the sample was fairly representative. The figure below summarizes descriptive statistics of the typical municipal IT head, in a South African municipal.

Table 3 – Municipal IT Head

| Profile Attribute (Municipal IT head) | Frequency |
|---|---|
| Not part of the executive team | 80.0% |
| Holds a minimum of an undergraduate qualification | 72.7% |
| African | 70.9% |
| Under 40 years | 87.3% |
| Male | 83.6% |
| Worked 1 to 5 years in a municipal | 55.77% |

Also, the table below summarizes characteristics of a typical municipality

| Table 4 – Municipal Profile | |
|---|---|
| Profile Attribute (Municipal) | Frequency |
| Poor municipals | 83.6% |
| Poor households | 94.5% |
| Minimum of 75 mobile phones per 100 individuals | 87.2% |

### 5.2.2. Frequencies for Likert-Scale Based Research Variables

Frequencies for Likert-scale based research variables are summarized in the table below.

Table 5 - Likert-Scale Based Research Variables

| Construct | Ratings | | |
|---|---|---|---|
| | 1 and 2 | 3 | 4 and 5 |
| IT security risks | 9.27% | 68.91% | 21.82% |
| Effectiveness of e-gov. enabling laws | 21.27% | 36.18% | 42.55% |
| IT's importance for municipal objectives | 7.45% | 33.28% | 59.27% |
| Use of IT by politicians | 46,00% | 48.55% | 5.45% |
| Adoption of e-government (resultant) | 42,00% | 43.28% | 14.72% |

### 5.3. Inferential Statistics

One way ANOVA tests were done between each profile item of municipalities' demographics and of their heads of IT, against the dependent variable on the perceived diffusion of e-government in municipalities; and correlation tests were also performed between each Likert-scale based research variable against the dependent variable.

### 5.3.1. One Way ANOVA

ANOVA tests between the perceived diffusion of e-government in municipalities and all ordinal items of the profile of the municipalities and their heads of IT did not show any relationship: urban or rural (F= 0.634), age (F = 0.554), educational level (F = 0.140), IT experience (F = 0.376), local government experience (F = 0.354), employment level (F = 0.551), teledensity (F = 0.937), municipality's service expenditure per annum (F = 0.589), annual household income (F = 0.661), and annual productivity index (F = 0.395).

ANOVA tests between the e-politician and all ordinal items of the profile of the municipalities and their heads of IT did not show any relationship: urban or rural (F=1.86), age (F = 0 .48), educational level (F = 1.05), IT experience (F = 1.43), local government experience (F = 0. 60), employment level (F = 0.81), teledensity (F = 1.08), municipality's service expenditure per annum (F = 0.72), annual household income (F = 0.30), and annual productivity index (F = 0.41).

### 5.3.2. Correlations

Table 6 – Correlations

|  |  | Profile | e. gov Security | e-gov Laws | Strategic Objectives | Politician IT use | IT Services |
|---|---|---|---|---|---|---|---|
| Profile | Pearson Corr. | 1 |  |  |  |  |  |
|  | 2-tailed Sig. |  |  |  |  |  |  |
|  | N | 54 |  |  |  |  |  |
| e-gov. Security | Pearson Corr. | -0.230 | 1 |  |  |  |  |
|  | 2-tailed Sig. | 0.095 |  |  |  |  |  |
|  | N | 54 | 55 |  |  |  |  |
| e-gov. laws | Pearson Corr. | -0.158 | 0.233 | 1 |  |  |  |
|  | 2-tailed Sig. | 0.255 | 0.088 |  |  |  |  |
|  | N | 54 | 55 | 55 |  |  |  |
| Strategic Objectives | Pearson Corr. | -0.047 | 0.138 | 0.187 | 1 |  |  |
|  | 2-tailed Sig. | 0.734 | 0.314 | 0.172 |  |  |  |
|  | N | 54 | 55 | 55 | 55 |  |  |
| Politician's IT use | Pearson Corr. | -0.046 | 0.406** | 0.378** | 0.156 | 1 |  |
|  | 2-tailed Sig. | 0.743 | 0.002 | 0.004 | 0.255 |  |  |
|  | N | 54 | 55 | 55 | 55 | 55 |  |
| IT Enabled Services | Pearson Corr. | 0.053 | 0.127 | 0.601** | 0.200 | 0.406** | 1 |
|  | 2-tailed Sig. | 0.706 | 0.355 | 0.000 | 0.143 | 0.002 |  |
|  | N | 54 | 55 | 55 | 55 | 55 | 55 |

## 6. FINDINGS AND DISCUSSIONS

Descriptive analysis of data collected through a survey of a sample of municipal IT heads finds a low rate of e-politicians (5.45%) and a corresponding low rate of e-service adoption (14.72%). This finding is supported by inferential findings that e-politician correlates with e-service adoption. The compatibility characteristic from the DoI theory of Rogers (2003), namely, that the degree to which the use of IT by politicians for constituency work is perceived as a better idea by very few politicians; conversely, very few municipals perceive the adoption of e-services as a better idea. Therefore, this study confirms knowledge from previous studies (Arduini *et al.*, 2010; Bwalya, 2009; Serrano-Cinca *et al.*, 2008; Weerakkody *et al.*, 2010). The relationship established between e-politician and e-service is depicted in figure 2, below, that summarizes overall findings.

However, descriptive analysis found that 87.2% of respondents believe that there are at least 75 mobile phones for 100 individuals. Thus, mobile phones are considered as a better idea by most municipal residents in accordance with Rogers (2003) DoI theory. This raises a question as to why most politicians are not using the mobile communications opportunity to advance their constituency work; as well, why are municipal IT heads not using the mobile telephone opportunity to provide municipal services online.

Laws facilitating e-services (e-laws) were rated by 42.55% of respondents as having an impact on e-service adoption. The descriptive finding on e-laws is strengthened by findings from inferential analysis that e-laws correlate with e-service. As well, inferential analysis found that e-laws correlate with e-politician, in accordance with compatibility characteristics from Rogers (2003) DoI theory. This finding about e-service, e-laws and e-politician confirms previous studies (Alpar and Sebastian, 2005; Basu, 2004). The finding suggests a relationship between e-laws, e-politician and e-services as depicted in figure 2, below, that summarizes overall results.

A puzzling 68.91% of respondents did not know whether e-security impacted e-service adoption, despite the fact that at least 72.7% of respondents said they are university graduates. Whereas e-service security (e-security) is a product of e-laws (Alpar and Sebastian, 2005; Basu, 2004), inferential analysis of this study found that e-security neither correlated with neither e-laws nor e-service. Instead, inferential analysis found that e-security correlates with e-politician, in accordance with compatibility characteristics from Rogers (2003) DoI theory. This finding suggests a relationship between e-security and e-politician is depicted in figure 2, below, that summarizes overall results.

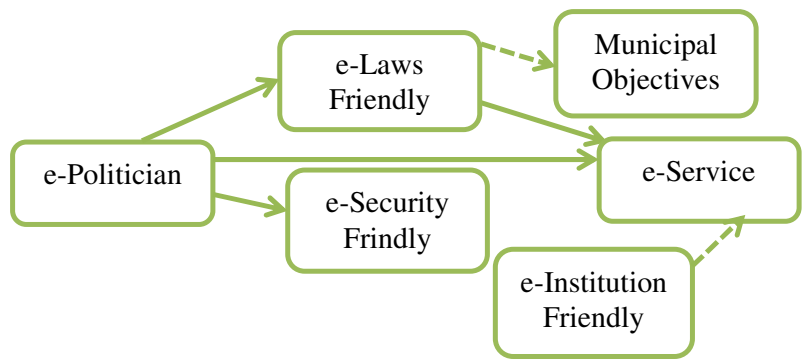

Figure 2 – Summarized Findings

Attributes of institutional arrangements form part of the profile of the municipal IT head and the municipal. Descriptive results of the profile, further to mobile teledensity findings that have already been presented with e-politician and e-service, show that 80% of respondents were not part of the executive team. Added to this, the study found that 42.8% of total municipals in South Africa did not have IT departments, in accordance with compatibility characteristics from Rogers (2003) DoI theory. The finding that 80% of respondents were not part of the executive team, begs a question from Li and Qiu (2010) assertions about who prepares and submits e-government motivations to elected municipal representatives for approval. Therefore, relationships are suggested by descriptive results between e-politician, e-Institution, and e-service as depicted in figure 2 that summarizes overall results.

The importance of IT for achieving municipal objectives was rated very high 59.27% by respondents, in accordance with compatibility characteristics from Rogers (2003) DoI theory. Nevertheless, this variable did not correlate with any other variable at all. This study suspects that the high rating was unduly influenced by the fact that a Mayor's performance contract is cascaded down to all municipal employees; and IT is purported as a tool of Municipal service delivery. Even so, a surprising 33.28% of respondents were neutral as to whether IT is important for the achievement of municipal objectives, and this may be explained by the possibility that non-municipal employees serve as IT heads in 42.8% of municipals. However, it was not the aim of this study to establish, amongst others, whether respondents were municipal employees or contracted IT providers. Therefore, relationships are suggested by descriptive results between municipal objectives and e-service as depicted in figure 2 that summarizes overall results.

There were other puzzling responses that could only be explained by the possibility that non-municipal employees served as IT heads: 48.55% of respondents did not know whether elected municipal representatives used IT for their political activities; 43.28% of respondents did not know whether municipal services were IT enabled; Alternatively, the tendency for neutral responses by municipal IT heads on IT matters may be resulting from the corresponding findings that 80.0% of the informants were not part of the executive team. An assertion of Carrizales (2008) and Kamal (2006) about the correlation of the presence of an IT department and e-service adoption is confirmed by the foregoing.

## 7. CONCLUSIONS AND FURTHER RESEARCH

This study aimed to determine whether municipal e-politicians have an impact on the adoption of municipal e-service in a developing country. The analysis of both descriptive and inferential analysis of data collected from municipal IT heads confirms that there is a relationship between e-politicians and e-service. Furthermore, this study found that the e-politician impacts e-laws that in turn impact e-service; e-politician impacts e-security; and e-politician impacts e-institution that in turn impacts e-service.

This conclusion has been summarised in figure 2 above. However, there were a noticeable percentage of respondents who tended to choose neutral responses. A further research is needed to clarify why most elected municipal representatives choose not to use IT for their constituency work, and why only few municipal services are IT enabled. Such further research should be located in the context of a finding of this study that 87.2% of respondents were aware that their developing country had at least 75 mobile telephones for 100 individuals.

**Author**

Ntjatji Gosebo serves as the Government Chief Information Officer in the Department of Public Service and Administration, South Africa. He studied Computer Science from New Jersey Institute of Technology, Newark, New Jersey, USA